\documentclass{PoS}

\title{BH masses in NLS1: the role of the broad-line region geometry}

\ShortTitle{BLR geometry and black hole masses in NLS1}

\author{\speaker{R. Decarli}\\
        Max-Planck Institut f\"ur Astronomie -- Heidelberg\\
        E-mail: \email{decarli@mpia.de}}

\author{M. Dotti\\
        Universit\`a degli Studi di Milano-Bicocca\\
        E-mail: \email{massimo.dotti@mib.infn.it}}
\author{F. Haardt\\
        Universit\`a degli Studi dell'Insubria -- Como\\
        E-mail: \email{francesco.haardt@uninsubria.it}}
\author{S. Zibetti\\
        DARK Cosmology center -- Copenhagen\\
        E-mail: \email{zibetti@dark-cosmology.dk}}

\abstract{Narrow Line Seyfert 1 galaxies (NLS1) are generally
believed to host ``under-massive'' black holes with respect to the
predictions from the host galaxy -- black hole mass scale relations.
Black hole masses in NLS1 are estimated from the continuum
luminosity and the width of broad emission lines. Here we show that
the ``mass deficit'' can be canceled out if we assume that the broad
line region (BLR) in type-1 AGN has a flat geometry, which is seen
face-on in NLS1. The detection of relativistic jets aligned along
the line of sight in a number of NLS1 supports this picture.
Moreover, a flat geometry of the BLR is also suggested by a general
trend of the mass deficit as a function of the line width observed
in other type-1 AGN, from quasars to BL Lac objects, and is
consistent with a simple extension of the Unified Model of AGN to
the BLR geometry. }

\FullConference{Narrow-Line Seyfert 1 Galaxies and their place in the Universe - NLS1,\\
        April 04-06, 2011\\
        Milan Italy}

\def\Mbh{$M_{\rm BH}$}
\def\Lhost{$L_{\rm host}$}
\def\Mhost{$M_{\rm host}$}

\def\edd{$L/L_{\rm Edd}$}

\def\Mgii{Mg\,{\sc ii}}
\def\Feii{Fe\,{\sc ii}}

\def\Oiii{[O\,{\sc iii}]}

\def\Ha{H$\alpha$}

\def\lsim{\mathrel{\rlap{\lower 3pt \hbox{$\sim$}} \raise 2.0pt \hbox{$<$}}}
\def\gsim{\mathrel{\rlap{\lower 3pt \hbox{$\sim$}} \raise 2.0pt \hbox{$>$}}}

\def\kms{km s$^{-1}$}

\begin{document}

\section{Introduction}
Narrow-line Seyfert 1 galaxies (NLS1) are a peculiar class of type-1
Active Galactic Nuclei (AGN) characterized by a modest width of
permitted broad lines (FWHM = 500--2000 \kms), weak \Oiii{} and
strong \Feii{} emission lines \cite{osterbrock85,talk_pogge}, strong
variability, and a soft X-ray excess \cite{boller03,grupe99}. They 
represent about 15\% of the whole population of Seyfert 1 galaxies 
\cite{williams02}, from optical spectroscopy classification.

In the context of the joint evolution of massive black holes (BHs)
and their host galaxies, the most important feature of NLS1 is that
they have, on average, lower \Mbh{} than expected from \Mbh{}--host
galaxy relations \cite{mathur01}, while \Mbh{} in normal broad line Seyfert
galaxies (hereafter, BLS1) are in fairly good agreement to the same
relation. On the other hand, AGN luminosities in NLS1 are comparable
to those of BLS1. Therefore, the ``\Mbh{} deficit'' directly yields
high Eddington ratios $L/L_{\rm Edd}$ in NLS1 \cite{talk_mathur}. 
Relatively modest BH masses are found from variability studies of
some NLS1 \cite{green93,hayashida00,mchardy06} and are sometimes associated with peculiar
properties of the host galaxy, e.g., pseudo-bulges \cite{talk_davies}
or enhanced star formation \cite{talk_sani}. In a BH--galaxy
co-evolution scenario, NLS1 could be still on their way to
reach the \Mbh{}--$\sigma_*$ relation, i.e., their (comparatively) small 
BHs are highly accreting in already formed bulges, thus moving towards
the BLS1 relation.

However, \Mbh{} in type-1 AGN are usually estimated through the so-called
virial paradigm, according to which the clouds in the broad line region
(BLR) orbit around the singularity following purely-gravitational dynamics.
Under this assumption, 
\begin{equation}\label{eq_virial}
M_{\rm BH} = G^{-1} \, R_{\rm BLR} \, v_{BLR}^2, 
\end{equation}
where
$G$ is the gravitational constant, and $R_{\rm BLR}$, $v_{BLR}$ are the
broad line region characteristic size and the velocity of broad-line
emitting clouds. $R_{\rm BLR}$ is found by means of the reverberation 
mapping technique \cite{blandford82}, or through the 
$R_{\rm BLR}$--luminosity relations \cite{kaspi00,kaspi05,kaspi07}.
The width of the broad emission lines, parametrized
e.g. by the full width at half maximum (FWHM), is used as a proxy for
$v_{\rm BLR}$, assuming a deprojection factor $f$:
\begin{equation}\label{eq_def_f}
v_{\rm BLR} = f \cdot {\rm FWHM}.
\end{equation}
The value of $f$ is generally unknown and depends on some assumptions 
about the geometry of the BLR. In the simple, isotropic scenario 
$f = \sqrt{3/4}$ \cite{netzer90}. If the BLR is rotationally supported
(see, e.g., \cite{wills86,bian04}),
an orientation dependence is introduced. If $\vartheta$ is the angle
between the rotation axis and the line of sight, the value of $f$
for a disk would be:
\begin{equation}\label{eq_def_fdisk}
f = \frac{1}{2} \left[\left(\frac{H}{R}\right)^2+\sin^2 \vartheta\right]^{-1/2}
\end{equation}
where $H$ and $R$ are the vertical and radial scale lengths of the disk. 
For infinitely thin disk,
$f$ ranges between $0.5$ and infinity as $\vartheta$ changes from 90
(edge-on) to 0 (pole-on) degrees. Assuming disks with finite thickness
would set an upper limit to $f$, $f_{\rm max}=R/(2\,H)$.

In order to measure $f$, two independent estimates of \Mbh{} are 
necessary, only one of which is based on equations 
\ref{eq_virial}--\ref{eq_def_fdisk}. The most common approach is to 
assume that the BH follows the \Mbh{}--host galaxy relations, therefore
the host properties (e.g., $\sigma_*$ or the bulge luminosity) can be
used as a proxy of \Mbh{} \cite{onken04,labita06,decarli08,graham11}
 (see however the caveats discussed in by Mathur \cite{talk_mathur}
and \cite{batcheldor10}).
All these studies found average values of $f$ exceeding unity, in
constrast with the simple expectations from the isotropic model.
In this picture, the observed small FWHM of NLS1 broad lines may be 
ascribed to a small viewing angle with respect to the disc axis, hence no 
evolutionary difference is required. 


Here we investigate this latter scenario from three different point 
of views: the demography of NLS1 with respect to BLS1 (section 
\ref{sec_nls1}); the geometric deprojection factor $f$ as derived in 
objects with small inclination angles $\vartheta$ (section \ref{sec_blazar}); 
and the comparison with the general quasar properties in terms of virial
black hole masses versus host galaxy properties (section \ref{sec_sdss}).
Conclusions are drawn in section \ref{sec_conclusions}.

\section{Breaking a cosmic conspiracy?}\label{sec_nls1}

\begin{figure}
\begin{center}
\includegraphics[width=0.49\textwidth]{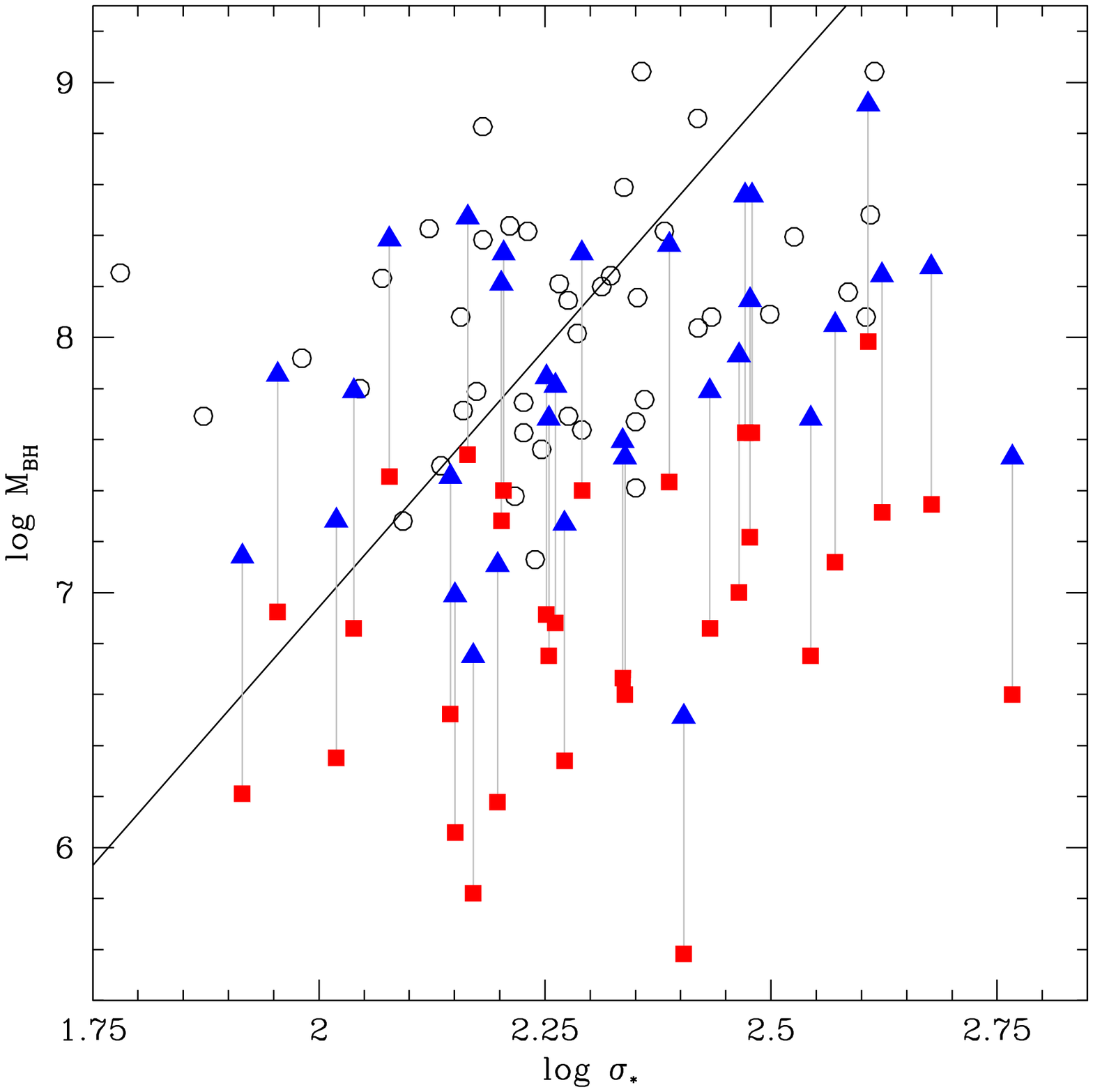}
\includegraphics[width=0.49\textwidth]{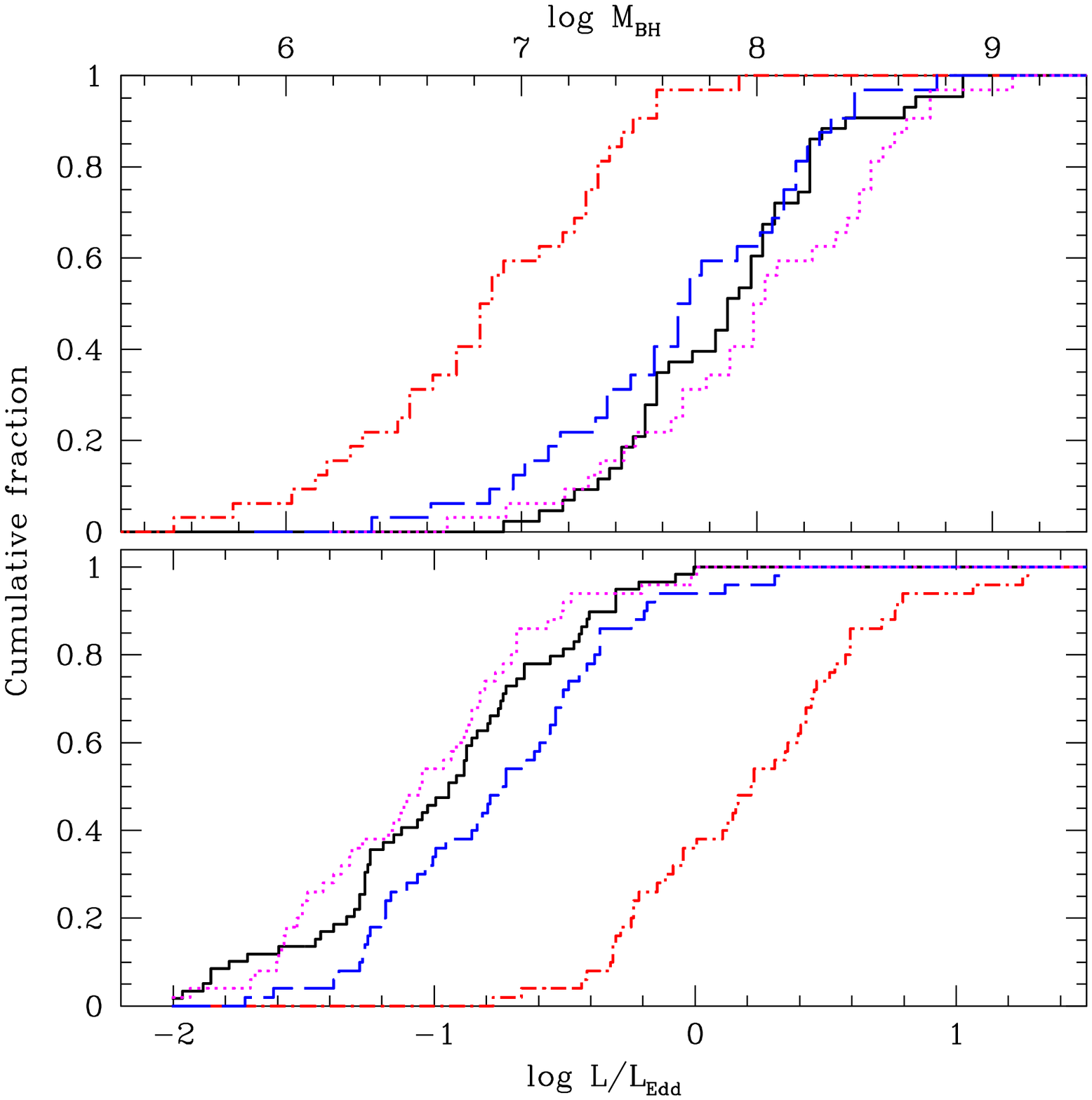}\\
\caption{\emph{Left panel}: The distribution
of BLS1 (empty, black circles) and NLS1 (squares and triangles) in the 
($\sigma_*$,\Mbh) plane. Data are taken from \cite{grupe04}.
Red squares refer to the isotropic case, 
while blue triangles refer to the disk-like BLR case with $H/R=0.1$.
After correcting for the BLR geometry, the two AGN classes
populate the same region of the plot. \emph{Right panel}: 
the cumulative distribution of \Mbh{} (\emph{top}) and
\edd{} (\emph{bottom}) of BLS1 (black, solid lines) and NLS1. 
The red, dot-dashed line is obtained assuming an 
isotropic BLR, with $f=\sqrt{3/4}$. The blue, dashed line and the 
magenta, dotted lines are obtained assuming a disk-like BLR with
$H/R$=0.1 and 0 respectively. The match with the \Mbh{} and \edd{}
distributions of normal BLS1 is remarkable. Thin disks with 
$0<H/R<0.1$ best fit the distributions. 
Figure adapted from \cite{nls1}.}\label{fig_nls1}
\end{center}
\end{figure}

The key point of the orientation scenario is the assumption that 
the BLR has a flat geometry, and that the relatively narrow width of 
permitted lines in
NLS1 galaxies is due to small inclination angles (pole-on views) 
rather than intrinsically narrow lines. This implies an underestimate
of the virial \Mbh{}, and consequently an overestimate of \edd{}. 
Under this hypothesis, we can estimate the average value of $f$
for NLS1 and BLS1 simply requiring that the solid angle of pole-on
systems and the one of non pole-on systems match the relative fractions of 
NLS1 and BLS1. According to the Unified Model of AGN, type-1 systems have
inclination angles ranging between 0 and $\vartheta_{\rm max}$ = 40--60 
degrees \cite{urry95}. We can define a $\vartheta_{\rm cut}$ so that 
objects with $\vartheta<\vartheta_{\rm cut}$ are labeled as NLS1, while
objects with $\vartheta_{\rm cut}<\vartheta<\vartheta_{\rm max}$ are 
considered BLS1. NLS1 constitute 15 \% of the BLS1 population, which 
yields $\vartheta_{\rm cut}\approx 15^\circ$, the exact value
depending on $H/R$. The corresponding average deprojection factors
are $\langle f \rangle_{\rm NLS1}\approx 3$ and 
$\langle f \rangle_{\rm BLS1}\approx 1$ (see \cite{nls1} for details).

Given these values, the position of BLS1 in the ($\sigma_*$,\Mbh{}) 
plane is practically unchanged with respect to the isotropic case
assumed, e.g., in \cite{grupe04}. On the other hand, BH masses in NLS1
galaxies are shifted $\approx 1$ dex upwards, thus matching the 
expectations from the \Mbh{}--$\sigma_*$ relation. Similarly, the
Eddington ratios are lowered by $\approx 1$ dex, matching the distribution
observed in normal BLS1. 

It is remarkable that with the only assumption that BLR is flat in 
\emph{all} type-1 Seyfert galaxies, and that they \emph{all} follow
the BH--host galaxy relations, we can break the ``cosmic conspiracy'' 
yielding relatively low \Mbh{} and similarly high \edd{} in AGN having
the same range of AGN luminosity and host galaxy properties (in this
case, $\sigma_*$). Once the orientation effect are taken into account,
the distributions of \Mbh{} and \edd{} in NLS1 and BLS1 are in good
agreement as well (see Figure \ref{fig_nls1}).

\section{The realm of small inclination angles}\label{sec_blazar}

\begin{figure}
\begin{center}
\includegraphics[width=0.49\textwidth]{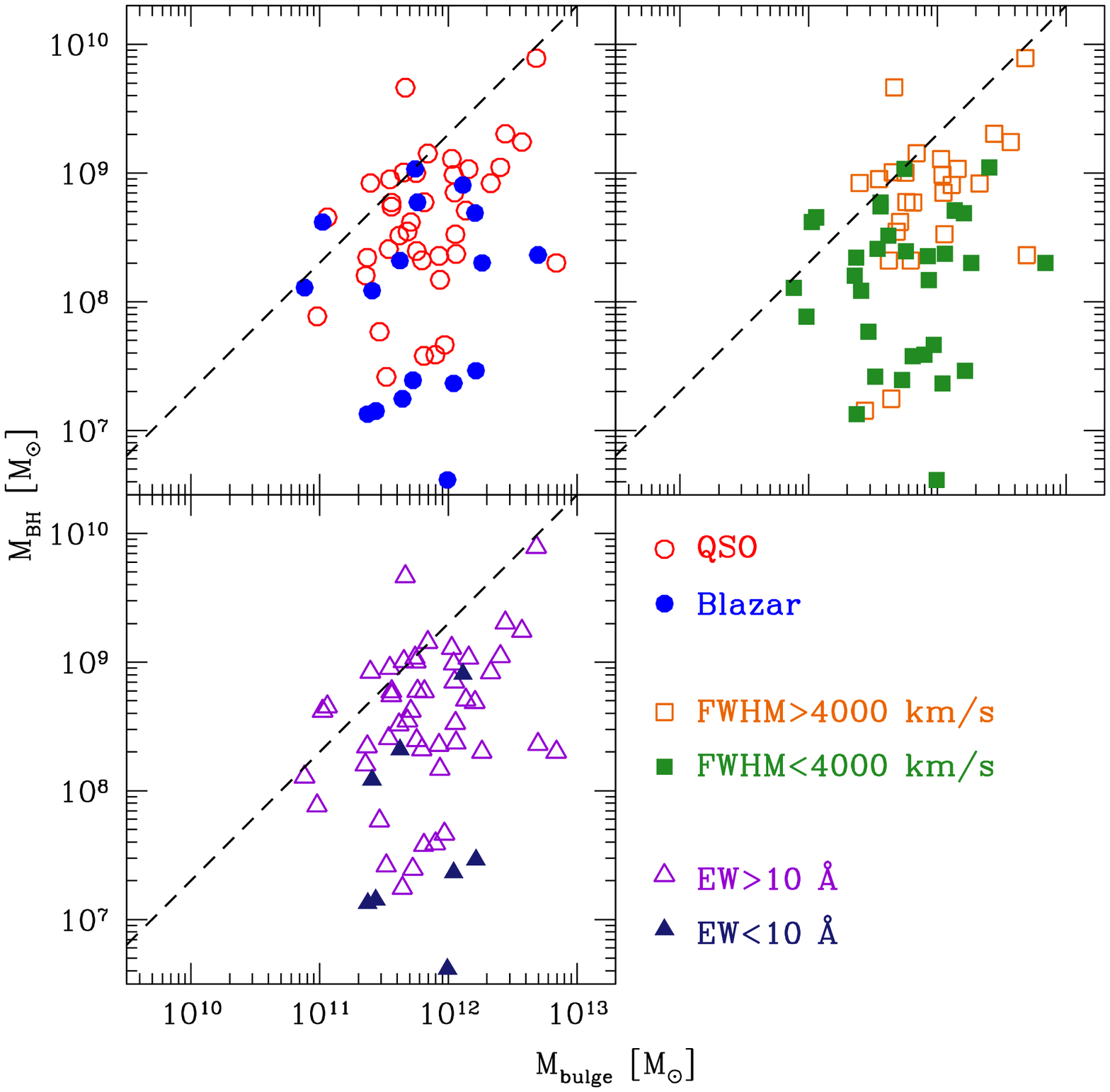}
\includegraphics[width=0.49\textwidth]{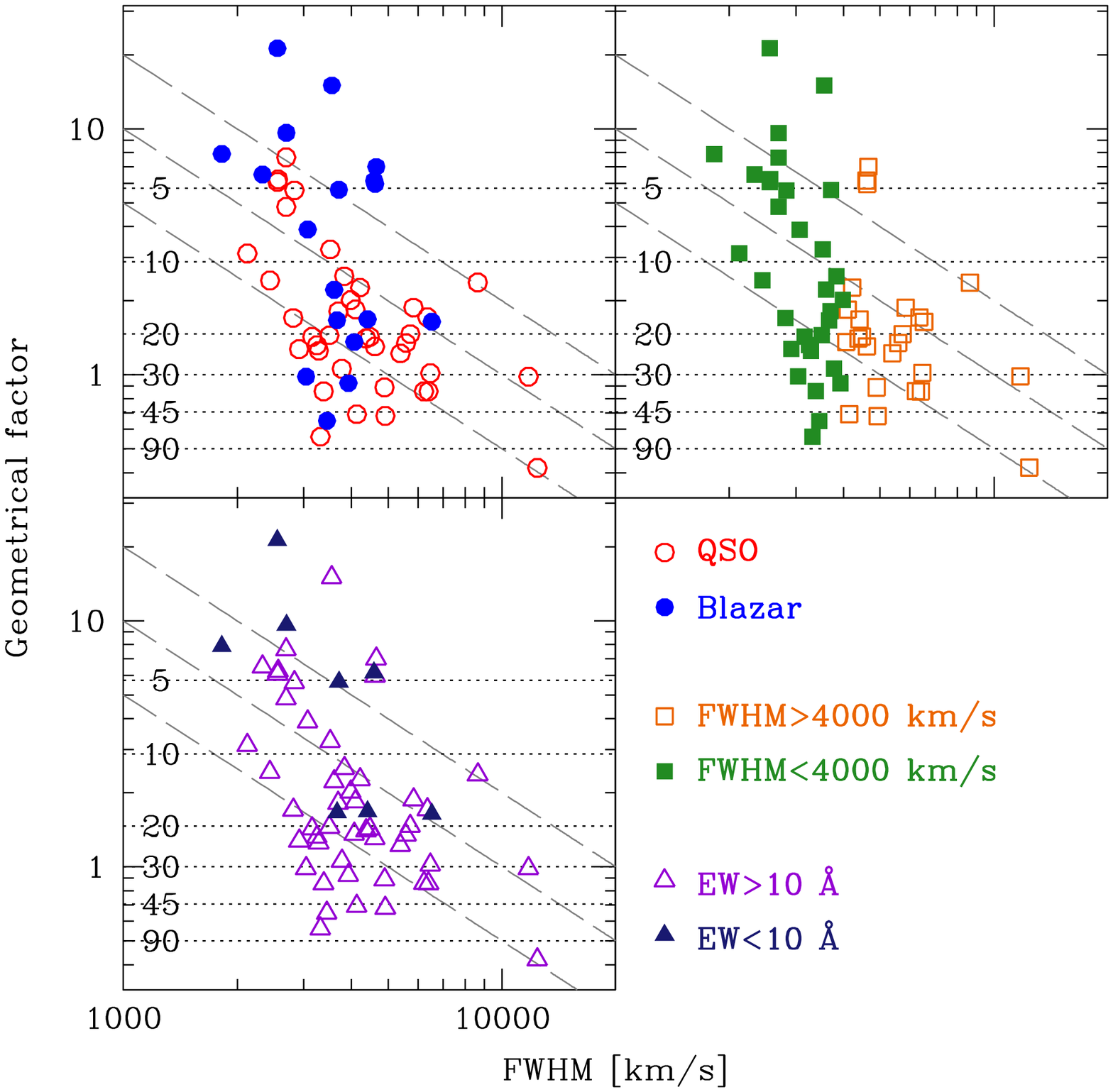}\\
\caption{\emph{Left panel}: the (\Mhost,\Mbh) plane in quasars and
blazars in our sample. Virial \Mbh{} with $f$=1 are plotted along the
$y$-axis. Blazars, and especially BL Lac objects (highlighted by
the EW$<$10 \AA{} cut) systematically deviate from the local
\Mbh{}=0.0015 \Mhost{} relation (dashed line).
\emph{Right panel}: the geometrical factor $f$ as a function of
the line width. A clear trend is apparent, with $f$ decreasing as
the line width increase. In particular, we note that objects with
relatively narrow lines ($<4000$ \kms) have systematically higher
$f$ values, consistently with the expectations for a thin disk-like 
BLR as seen from various inclination angles (5, 10, 20, 30, 45 and 90
degrees, horizontal dotted lines in the plot; see equation 1.3).
BL Lac objects, known to have small $\vartheta$, populate the top-left
corner of this plot, with relatively narrow lines and high $f$,
consistently with the finding for NLS1 described in section 2. Figure
adapted from \cite{decarli11}.
}\label{fig_blazar}
\end{center}
\end{figure}

In order to test the orientation scenario, one would ideally have
an independent way to measure the inclination. We therefore focus now on a
class of type-1 AGN in which inclination angles are known to be small.
Blazars, and BL Lac objects among them, are characterized by strong 
continuum luminosity in the optical and radio wavelengths, associated to 
the synchrotron emission from a relativistic jet almost aligned with the 
line of sight, i.e., $\vartheta\lsim10^{\circ}$ \cite{ghisellini93,urry95}.
Therefore, we expect BL Lac objects to show substantially higher $f$ values
than the typical quasars.

We select a sample of quasars and blazars requiring that
the host galaxy has been resolved and it's elliptical (so that 
$L_{\rm bulge}\approx L_{\rm host}$), and that broad \Mgii{} or \Ha{} 
lines have been observed in their optical spectra. Quasars are
taken from \cite{decarli10a,decarli10b}, while blazars are
collected mainly from \cite{scarpa00,kotilainen98,nilsson03}.
The full list of sources and references for the data is given in
\cite{decarli11}. The final sample consists of 18 blazars (including 
8 BL Lac objects) and 39 quasars.

Virial BH masses are derived for each object using the broad line 
width (FWHM) and luminosity ($L_{\rm line}$) used as a proxy of the 
accretion disk continuum luminosity\footnote{A fraction of the 
photoionizing radiation illuminating the BLR in blazars could in principle
come from the jet, rather than from the accretion disk. However,
line luminosities are found to be independent of the jet-associated continuum
\cite{corbett00}.}. The host galaxy stellar mass (\Mhost) 
is derived from the rest-frame $R$-band luminosity of the host galaxy, 
after applying $k$-correction and assuming the mass-to-light ratio of a 
single stellar population formed at high redshift and passively evolving 
(see \cite{decarli10b,decarli11}). Finally, we adopt the \Mbh{}/\Mhost{}
ratio and its redshift dependence as found in \cite{decarli10b} in 
order to compute $f$.

Blazars, and in particular BL Lac objects, show important (up to 2 dex!) 
deviations with respect to the local \Mbh{}/\Mhost{} relation, if the
geometrical factor is not taken into account (see Figure \ref{fig_blazar},
\emph{left}). Most importantly, the geometrical factor $f$ shows a clear
dependence on the line width (see Figure \ref{fig_blazar}, \emph{right}),
so that objects with modest line width have high ($>3$) values of $f$.
It is remarkable that blazars, and specifically BL Lac objects, populate
the top-left corner of the (FWHM,$f$) plane, with relatively narrow lines
and high deprojection factors, consistently with the disk-like picture
of the BLR. This trend is in bitter disagreement with the isotropic
model of the BLR, according to which the geometrical factor is constant
and smaller than 1. It is interesting to note that some objects have
very high $f$ values, thus supporting the idea of a geometrically thin
($H/R<0.1$) disk.

\section{Another look at the \Mbh{}--\Mhost{} relation in quasars}\label{sec_sdss}

\begin{figure}
\begin{center}
\includegraphics[width=0.49\textwidth]{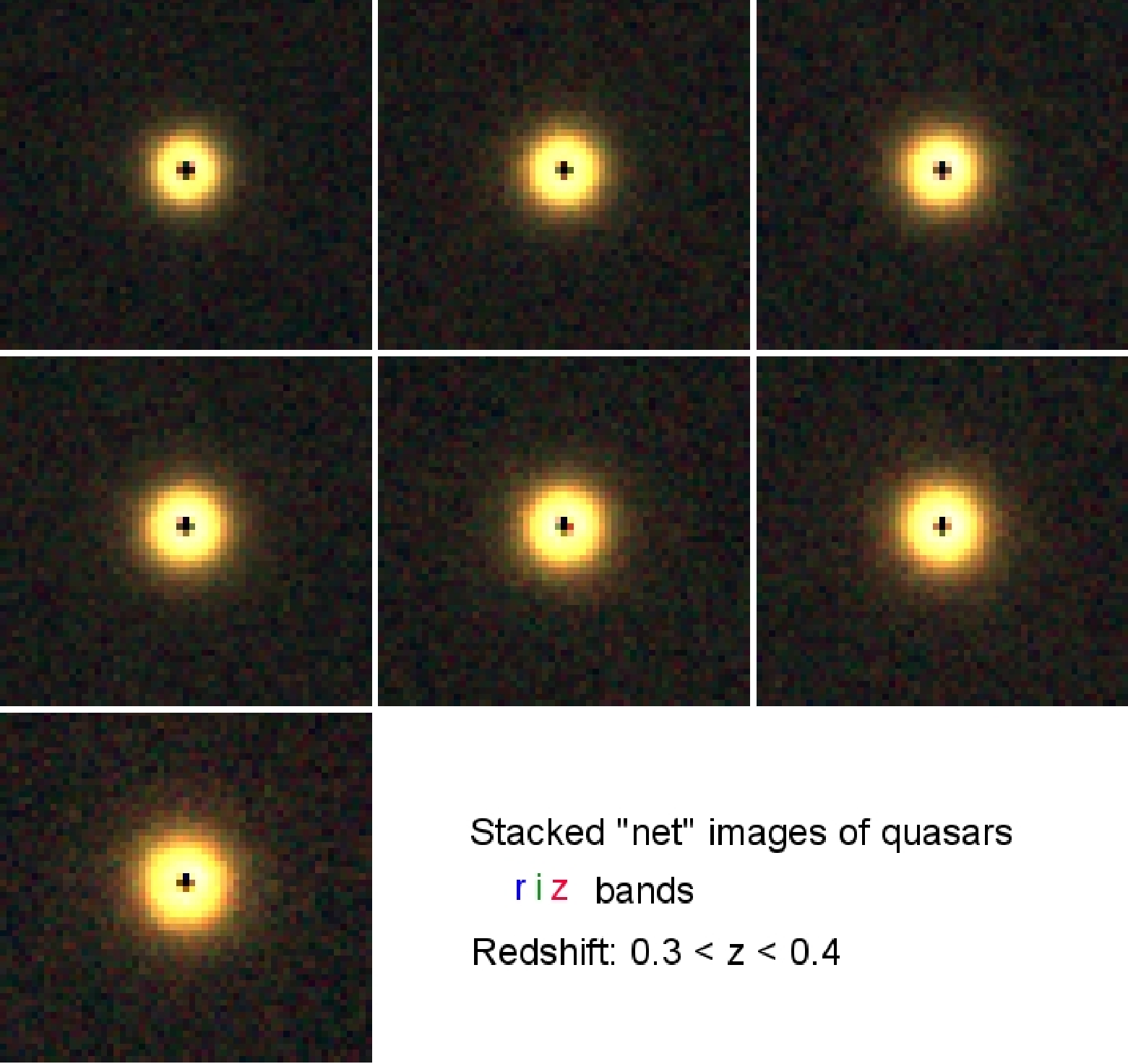}
\includegraphics[width=0.49\textwidth]{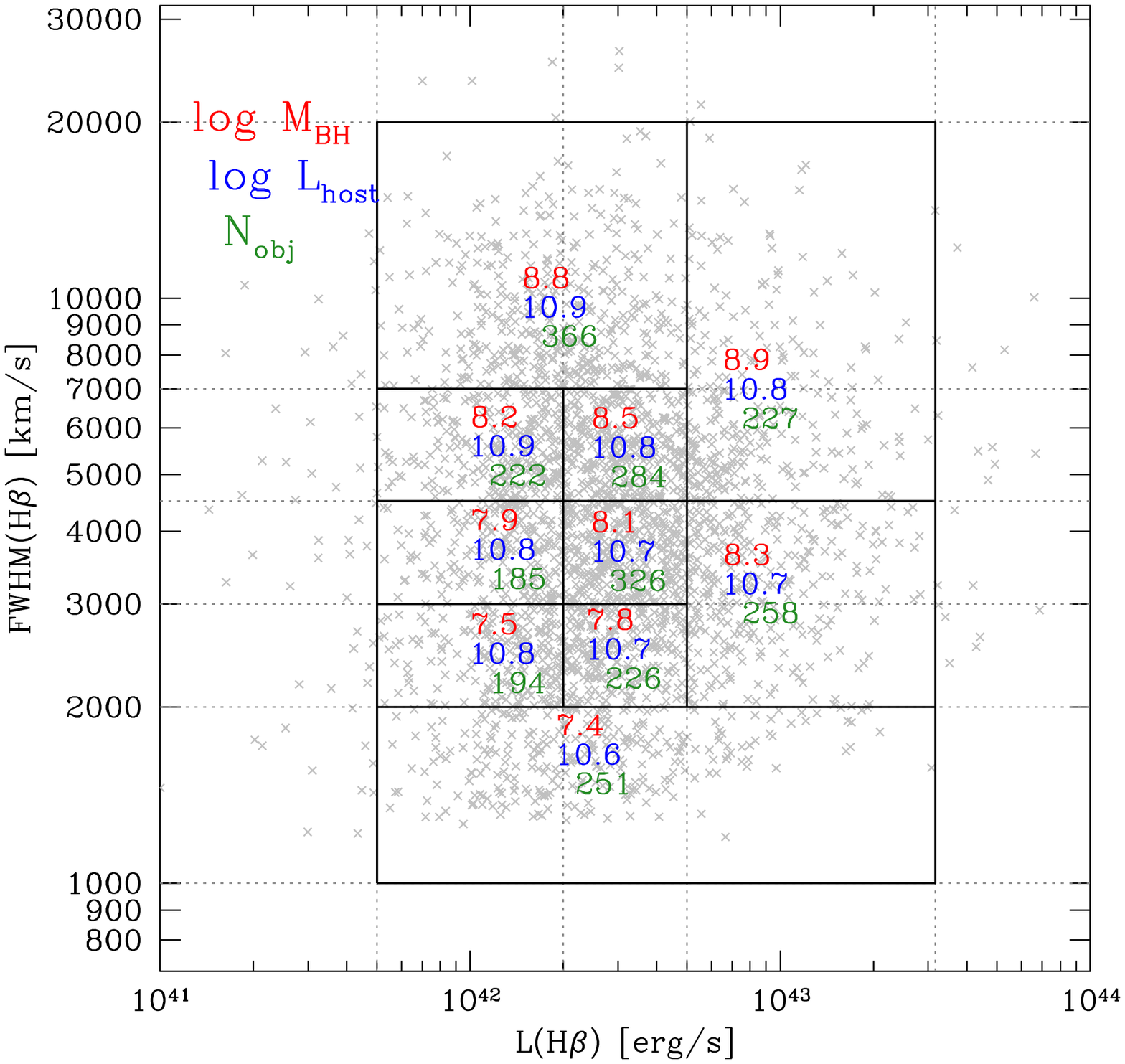}\\
\caption{\emph{Left panel}: stacked ``net'' images of
quasars at $0.3<z<0.4$ in various \Mbh{} bins. The RGB images
are created as a RGC composite of the SDSS $zir$ stacks. Net images 
are obtained by subtracting from the quasar image the image
of a bright field star, scaled to match the cental brightness
of the target. The extended emission of the host galaxies is
clearly detected after the stacking analysis.
\emph{Right panel}: the ($L_{\rm line}$, FWHM) plane for SDSS 
quasars at $0.3<z<0.4$ (grey crosses). Data are taken from 
\cite{shen10}. We rebinned the sample in blocks of comparable
FWHM and $L_{\rm line}$ (thick black lines), and computed
the average \Mbh{} in each bin (values in red). As expected,
\Mbh{} increases towards to top-right corner of the plot.
We then performed our stacking analysis on each bin. The resulting
host galaxy luminosity is shown in blue. While the FWHM increases
by a factor $\sim 5$ from bottom to top (\Mbh{} increases by
a factor $\sim 20$), only a small increase in terms of \Lhost{}
is reported (a factor $\sim 2$), suggesting that the line
width, taken without any correction for the $f$--FWHM dependence,
is not a good tracer of \Mbh{}.
}\label{fig_stack}
\end{center}
\end{figure}

The study presented in section \ref{sec_blazar} is strongly limited
by the modest size of quasar and blazar samples with resolved host
galaxy available in the literature. This is because the extended component
associated to the galaxy is outshone by the bright nuclei, and optimal
angular resolutions are required to probe the host galaxy.
However, it is possible to derive average properties of quasar host
galaxies from stacking several hundred images of quasars observed with
non-optimal seeing conditions, e.g., from the Sloan Digital Sky Survey.
The power of this approach is shown in Figure \ref{fig_stack}, \emph{left},
where we show the ``net'' images of stacked quasars at $0.3<z<0.4$ binned
as a function of their virial BH mass. Net images are obtained by 
subtracting from the quasar image the image of a bright star in the field,
scaled to match the central surface brightness of the target. Since 
host galaxies are rarely resolved in typical SDSS data, individual 
net images are noise-dominated. However, once we stack few hundred of sources,
the extended light from the host galaxy is apparent. 

We performed our stacking analysis on SDSS quasars ranging between 
$z=0.3$ and $z=1.3$ (in order to have significant host galaxy emission
in the SDSS filters), and binning as a function of spectral properties
of the quasars, e.g., \Mbh{}, FWHM or luminosity of some broad line, etc.
Then we modeled the stacked image of the quasars using a Sersic profile
for the host galaxy and the PSF model derived from the stacked image of
field stars. The resulting magnitudes in 4 bands (SDSS $griz$) are then
compared to get color information and to give constraints on the stellar 
population of the host galaxies. This allows us to probe the dependence 
of the average \Lhost{} and \Mhost{} in each bin as a function of
the quasar properties and the redshift. Details will be presented in
\cite{dz11}.

In Figure \ref{fig_stack}, \emph{right} we plot the $0.3<z<0.4$ quasars
from \cite{shen10} in the ($L_{\rm line}$, FWHM) plane. We binned the
plane in terms of line width and luminosity, and computed the average
virial \Mbh{} in each bin. Then, we perfomed our stacking analysis, and
evaluated the average host galaxy luminosity. Surprisingly, while the 
FWHM increases by a factor $\sim5$ from the bottom to the top of the plot,
and the average \Mbh{} increases by a factor $\sim20$, the host galaxy
luminosity changes only by a factor $\sim2$ 
from the bottom to the top of the diagram. This result is in disagreement
with the observed \Mbh{}--\Mhost{} relation. If, as shown before, the 
geometrical factor shows a dependence on the FWHM, then the FWHM could
have only a limited power in predicting the actual $v_{\rm BLR}$. This
could explain the lack of correlation between \Mbh{} and host luminosities
from the stacking analysis, as bins are defined on virial \Mbh{} estimates.

\section{Conclusions}\label{sec_conclusions}
We showed that the relatively small \Mbh{} and high \edd{} values 
observed in NLS1 may be fully attributed the orientation of the BLR 
with respect to the line of sight, once a flat BLR geometry is assumed.
This argument can simultaneously break the ``cosmic conspiracy'' in NLS1,
explain the mismatch between virial \Mbh{} and \Mbh{} obtained 
from the host galaxy properties in blazars, and accounts for
the lack of any significant correlation between virial \Mbh{} and host
galaxy luminosities in quasars.


\begin{thebibliography}{99}
\bibitem{osterbrock85} Osterbrock D.E., Pogge R.W., 1985, ApJ, 297, 166
\bibitem{talk_pogge} R.~W. Pogge: \emph{A quarter century of Narrow-Line Sefert 1s}. In: \emph{Proceedings of the Workshop Narrow-Line Seyfert 1 Galaxies and Their Place in the Universe}, eds L. Foschini, M. Colpi, L. Gallo, D. Grupe, S. Komossa, K. Leighly, \& S. Mathur. \emph{Proceedings of Science (NLS1) 002} (2011).
\bibitem{boller03} Boller T., Tanaka Y., Fabian A., Brandt W.N., Gallo L., Anabuki N., Haba Y., Waughan S., 2003, MNRAS, 343, L89
\bibitem{grupe99} Grupe D., Beuermann K., Mannheim K. Thomas H.C., 1999, A\&A, 350, 805
\bibitem{williams02} Williams R.J., Pogge R.W., Mathur S., 2002, AJ, 124, 3042
\bibitem{mathur01} Mathur S., et al. 2001, New Astronomy, 6, 321
\bibitem{talk_mathur} S. Mathur: \emph{Host galaxies of NLS1s}. In: \emph{Proceedings of the Workshop Narrow-Line Seyfert 1 Galaxies and Their Place in the Universe}, eds L. Foschini, M. Colpi, L. Gallo, D. Grupe, S. Komossa, K. Leighly, \& S. Mathur. \emph{Proceedings of Science (NLS1) 002} (2011).
\bibitem{green93} Green A.R., McHardy I.M., Lehto H.J., 1993, MNRAS, 265, 664
\bibitem{hayashida00} Hayashida K., 2000, NewAR, 44, 419
\bibitem{mchardy06} McHardy I.M., Koerding E., Knigge C., Uttley P., Fender R.P., 2006, Nature, 444, 730
\bibitem{talk_davies} R. Davies: \emph{Cosmic Evolution of NLS1 and the Growth of their Black Holes}. In: \emph{Proceedings of the Workshop Narrow-Line Seyfert 1 Galaxies and Their Place in the Universe}, eds L. Foschini, M. Colpi, L. Gallo, D. Grupe, S. Komossa, K. Leighly, \& S. Mathur. \emph{Proceedings of Science (NLS1) 002} (2011).
\bibitem{talk_sani} E. Sani: \emph{Enhanced star formation in Narrow-Line Seyfert 1 galaxies}. In: \emph{Proceedings of the Workshop Narrow-Line Seyfert 1 Galaxies and Their Place in the Universe}, eds L. Foschini, M. Colpi, L. Gallo, D. Grupe, S. Komossa, K. Leighly, \& S. Mathur. \emph{Proceedings of Science (NLS1) 002} (2011).
\bibitem{netzer90} H. Netzer, 1990, in Blandford R.D., Netzer H., Woltjer L., eds, Active Galactic Nuclei, Springer, Berlin , p. 137
\bibitem{wills86} Wills B.J., Browne  I.W.A., 1986, ApJ, 302, 56
\bibitem{bian04} Bian W. \& Zhao Y., 2004, MNRAS, 352, 823
\bibitem{blandford82} Blandford R.D., McKee C.F., 1982, ApJ, 255, 419
\bibitem{kaspi00} Kaspi S., Smith P.S., Netzer H., Maoz D., Jannuzi B.T., Giveon U., 2000, ApJ, 533, 631
\bibitem{kaspi05} Kaspi S., Maoz D., Netzer H., Peterson B.M., Vestergaard M., Jannuzi B.T., 2005, ApJ, 629, 61
\bibitem{kaspi07} Kaspi S., Brandt W.N., Maoz D., Netzer H., Schneider D.P., Shemmer O., 2007, ApJ, 659, 997
\bibitem{onken04} Onken C.A., Ferrarese L., Merritt D., Peterson B.M., Pogge R.W., Vestergaard M., Wandel A., 2004, ApJ, 615, 645
\bibitem{labita06} Labita M., Treves A., Falomo R., Uslenghi M., 2006, MNRAS, 373, 551
\bibitem{decarli08} Decarli R., Labita M., Treves A., Falomo R., 2008a, MNRAS, 387, 1237
\bibitem{graham11} Graham A.W., Onken C.A., Athanassoula E., Combes F., 2011, MNRAS, 412, 2211
\bibitem{batcheldor10} Batcheldor D., 2010, ApJ, 711L, 108
\bibitem{urry95} Urry C.M., Padovani P., 1995, PASP, 107, 803
\bibitem{grupe04} Grupe D., Mathur S., 2004, ApJ, 606L, 41
\bibitem{nls1} Decarli R., Dotti M., Fontana M., Haardt F., 2008b, MNRAS Letters, 386, 15
\bibitem{ghisellini93} Ghisellini G., Padovani P., Celotti A., Maraschi L., 1993, ApJ, 407, 65
\bibitem{decarli10a} Decarli R., Falomo R., Treves A., Kotilainen J.K., Labita M., Scarpa R., 2010a, MNRAS, 402, 2441
\bibitem{decarli10b} Decarli R., Falomo R., Treves A., Labita M., Kotilainen J.K., Scarpa R., 2010b, MNRAS, 402, 2453
\bibitem{scarpa00} Scarpa R., et al., 2000, ApJ, 532, 740
\bibitem{kotilainen98} Kotilainen J.K., Falomo R., \& Scarpa R., 1998, A\&A, 332, 503
\bibitem{nilsson03} Nilsson K., Pursimo T., Heidt J., Takalo L.O., Sillanp\"a\" A., Brinkmann W., 2003, A\&A, 400, 95
\bibitem{decarli11} Decarli R., Dotti M., Treves A., 2011, MNRAS, 413, 39
\bibitem{corbett00} Corbett E.A., Robinson A., Axon D.J., Hough J.H., 2000, MNRAS, 311, 485
\bibitem{dz11} Decarli R., Zibetti S., 2011, in preparation
\bibitem{shen10} Shen Y., Richards G.T., Strauss M.A., Hall P.B., Schneider D.P., Snedden S., Bizyaev D., Brewington H., et al., 2010, arXiv:1006.5178

%

%

\end{thebibliography}
\end{document}